\documentclass[a4paper,12pt]{article}
\usepackage{amsmath,amsfonts,amssymb,amsthm,amstext}
\usepackage[cp1251]{inputenc}
\usepackage[T2A]{fontenc}
\usepackage[russian,english]{babel}
\newcommand{\algdef}[4]{%
\subsubsection*
{\bfseries{\itshape Algorithm #1.}\\
#2.}%
\emph{Input:} #3.\\%
\emph{Output:} #4.\\%
\emph{Description:}%
}

\newcommand{\algdefo}[5]{%
\subsubsection*
{\bfseries{\itshape Algorithm #1.}\\
#2.}%
\emph{Input:} #3.\\%
\emph{Output:} #4.\\%
\emph{Oracles:} #5.\\%
\emph{Description:}%
}

\newcommand{\algdefpo}[6]{%
\subsubsection*
{\bfseries{\itshape Algorithm #1.}\\
#2.}%
\emph{Input:} #3.\\%
\emph{Output:} #4.\\%
\emph{Parameters:} #5.\\%
\emph{Oracles:} #6.\\%
\emph{Description:}%
}

\newcommand{\classnameindef}[1]{%
{\normalfont\bf #1}}

\newcommand{\classname}[1]{%
{\bf #1}}

\newcommand{\setnameinform}[1]{%
\mathbf{#1}}

\newcommand{\algname}[1]{%
{\normalfont\itshape #1}}

\newcommand{\algnameindef}[1]{%
{\normalfont\itshape #1}}

\begin{document}

\title{
Time- and space-efficient evaluation of the complex\\
exponential function using series expansion}

\author{Sergey~V.~Yakhontov\\
Ph.D. in Computer Science\\
Faculty of Mathematics and Mechanics\\
Saint Petersburg State University\\
Russian Federation\\
SergeyV.Yakhontov@gmail.com\\
Phone: +7-911-966-84-30\\
14-Aug-2012}
\date{}

\maketitle

\newtheorem{notation}{Notation}
\newtheorem{definition}{Definition}
\newtheorem{proposition}{Proposition}
\newtheorem{lemma}{Lemma}
\newtheorem{theorem}{Theorem}
\newtheorem{corollary}{Corollary}

\abstract {
An algorithm for the evaluation
of the complex exponential function is proposed which is
{\bf quasi-linear in time and linear in space}. This algorithm
is based on a modified binary splitting method for the hypergeometric
series and a modified Karatsuba method for the fast evaluation of
the exponential function. The time complexity of this algorithm
is equal to that of the ordinary algorithm for the
evaluation of the exponential function based on the series expansion: $O(M(n)\log(n)^2)$.
}


\paragraph{1. Introduction.}
In this paper we introduce a measure of the space complexity of calculations
on a Schonhage machine \cite{Schonhage94} and give an upper bound for the
time and space complexity of a proposed algorithm for the computation
of the exponential function of a complex argument in each area $|z|\le 2^p$
where $p$ is a natural number, $p\geq 0$, on the Schonhage machine.

The Schonhage machine is in fact an ordinary computer. Therefore,
we describe an algorithm which is quasi-linear in time and linear in space
on an ordinary computer. Hence, the phrase 
`fast algorithm' refers to evaluations on a Schonhage machine.
In particular, the estimate $O(n\log(n)\log\log(n))$ refers to this
machine \cite{Schonhage94}.

Basic information on dyadic rational numbers and constructive real
numbers and functions can be found in \cite{Ko91}. The notation
\classname{Sch(FQLINTIME//LINSPACE)} will be used for the class of algorithms
which are quasi-linear in time and linear in space on a Schonhage machine.
Quasi-linear means that the complexity function is bounded by
$O(n\log(n)^k)$ for some $k$.

From now on, $n$ will denote the length of the record of accuracy
$2^{-n}$ of dyadic rational approximations; $x$ will be used for a real
argument, $z$ wiil be used for a complex argument. We will use $\log(k)$ for
logarithms base 2.

The basic subject of interest is the algorithms for the evaluation of
elementary functions based on series expansions, as such algorithms are
important for practical computer science due to the relative simplicity
of their implementation.

Algorithms for the fast evaluation of the exponential function and some other
elementary functions with a time complexity of $O(M(n)\log(n)^{2})$
are offered in \cite{Karatsuba91} (where $M(n)$ denotes the complexity of
multiplication of $n$-bit integers); the space used by these algorithms is
bounded by $O(n\log(n))$, as will be shown below. In \cite{Haible98}, algorithms for the calculation of elementary functions, based on
Taylor series, which are quasi-linear in
time are considered; the space used by the algorithms from
\cite{Haible98} is bounded by a quasi-linear function as the classical
binary splitting method uses $O(n\log(n))$ space for the intermediate results.

That is, elementary functions are computable using quasi-linear
time and quasi-linear space. There is a question: whether it is possible
to evaluate elementary functions using quasi-linear in time
and linear in space algorithms based on series expansions?

This paper shows that the answer to this question for the
complex exponential function and some other complex elementary
functions is Yes. We use the combination of two algorithms
for the construction of an algorithm of complexity class
\classname{Sch(FQLINTIME//LINSPACE)} for the evaluation of the exponential
function: a modified binary splitting method for the evaluation of
the hypergeometric series, and a modified Karatsuba method
\cite{Karatsuba91} for the fast evaluation of the exponential
function.

Note that the residual sum of
the series \eqref{Eq:ExpGammaNu} satisfies the following inequality:
\begin{equation*}
|R_{\nu}(r)|<C2^{-(r\log(r)+m)};
\end{equation*}
here $r=m 2^{-\nu+1}$. Therefore, $|R_{\nu}(r)|=O(2^{-m\log(m)})$
doesn't hold, and we cannot get the result about \classname{FLINSPACE}
computability of $\exp(x)$ from this equation.


\paragraph{2. Description of the computation model
(machine Schonhage).}
This machine, introduced in \cite{Schonhage94}, operates on symbols of the
alphabet $\Sigma=\{0,1,\ldots,2^{\lambda-1}\}$ and sequences of such symbols
(we can take, for example, a constant $\lambda$ equal to 32).
The machine consists of arrays $T_0,\dots, T_{\tau}$ to read and write
symbols from $\Sigma$, registers $A$, $B$, $C$, $M$ for arithmetic
operations, and a control unit CPU. The arrays are infinite in both
directions. For each array there is a pointer $p_i$ to the current
symbol written in the array. The record $<p+j>$ means the symbol
referenced by pointer $p+j$. There is also an additional register
$Y$ which is a pointer to the current operation, and an optional
array $S$ which serves as the stack of recursive calls; $S$ is infinite
in one direction. A bit register $E$ acts as an overflow register
for arithmetic operations.

A program for Schonhage consists of several modules written in
the language TPAL, which is similar to an assembly language for a RISC
processor. In TPAL there are commands for loading a symbol written
in an array into a register, for reading a symbol from
a register and writing it to an array, for increasing and
decreasing the content of a register, the shift command,
the call and return from a procedure commands, the jump to a label
command, and the conditional jump command. Integers on which the
machine operates are encoded as symbol sequences in the alphabet
$\Sigma$:
\begin{equation*}
a=a_0+a_1 2^{\lambda}+a_2(2^{\lambda})^2+\ldots+
a_{k-1}(2^{\lambda})^{k-1},\quad 0\le a_i\le 2^{\lambda}-1.
\end{equation*}
The sign bit is written in the symbol which is before the senior
symbol $a_{k-1}$.

Schonhage can call procedures and perform recursive calls. After the
return from a procedure the memory occupied by the parameters
and local variables is released.

The  time computational complexity of an algorithm on Schonhage is
defined as the number of instructions in the language TPAL. 
Arithmetic operations on symbols and calls of procedures are
counted as a constant number of steps. The memory used in an array
during the calculation is defined as the maximum of the number of
array elements involved in the calculation.

As constructive functions are functions that compute approximations
of functions using approximations of arguments, we define the oracle
machine Schonhage. This machine has some oracle functions that compute
approximations of arguments; the machine calculates approximations
of a function using these approximations of arguments. A request to
an oracle is written in array $T_0$ as the record of an accuracy of
the computation; approximations of arguments are recorded in array
$T_0$ too. A query to an oracle is treated as one operation in the
time computational complexity of the oracle machine Schonhage.

\begin{definition}
The space computational complexity of an algorithm on
the oracle machine Schonhage is defined as the sum of the memory used
for all the arrays plus the maximum of the memory used for the stack.
\end{definition}


\paragraph{3. Constructive complex numbers and functions.}
A complex number $z'$ such that $|z-z'|\le 2^{-n}$
is called
an approximation of the complex number $z$ with accuracy $2^{-n}$.

Suppose that there are complex numbers $\omega=x+iy$, $\omega'=x'+iy'$
such that $|x-x'|\le 2^{-(n +1)}$ and $|y-y'|\le 2^{-(n +1)}$.
Then
\begin{equation}
|\omega-\omega'|=\sqrt{(x-x')^2+(y-y')^2}\le\sqrt{2\cdot 2^{-2(n+1)}}<2^{-n}.
\label{Eq:ComplNumberApprox}
\end{equation}
That is, to calculate an approximation of complex number $\omega$ with
accuracy $2^{-n}$ it is sufficient to calculate an approximation of the real
and imaginary parts of the complex number with accuracy $2^{-(n +1)}$.

We say that a sequence
$\phi:\setnameinform{N}\rightarrow\setnameinform{D}\times\setnameinform{D}$,
where $\phi(n)=(\phi_x(n),\phi_y(n))$, $\setnameinform{D}$ is the set of dyadic
rational numbers, converges dyadic-rationally to the complex number $z$ if
for any $n\in\setnameinform{N}$ the following holds: $prec(\phi_x(n))=n+2$,
$prec(\phi_y(n))=n+2$, and $|z(n)-z|\leq 2^{-n}$,
where $z(n)=\phi_x(n)+i\phi_y(n)$. The set of all functions $\phi$
which converge dyadic-rationally to a number $z$ is denoted by
$CF_z$. A complex number $z$ is called a $CF$ constructive complex number
if $CF_z$ contains a computable function $\phi$.

\begin{definition}
A complex number $z\in\mathbb{C}$ is called
a \classname{Sch(FQLINTIME//LINSPACE)} constructive complex number
if there exists a function $\phi\in CF_z$ which belongs to
the class \classname{Sch(FQLINTIME//LINSPACE)}.
\end{definition}

Let $f$ be a function $f(z):A\to\mathbb{C}$,
where $A=\{z\in\mathbb{C}:|z|\le R\}$ is an area in the set of
complex numbers.

\begin{definition}
The function $f(z)$ is called a \classname{Sch(FQLINTIME//LINSPACE)}
constructive complex function in the area $A$ if for any $z$ from
this area there is a function $\psi$ from $CF_{f(z)}$
which belongs to the class \classname{Sch(FQLINTIME//LINSPACE)}.
\end{definition}

Note that to calculate the values of a constructive complex function
of the argument $z=x+iy$ we need to specify functions $u(x,y)=Re(f(z))$ and
$v(x,y)=Im(f(z))$, and in order to calculate these functions we need
to have two oracle functions that correspond to the real and imaginary
parts of the argument.


\paragraph{4. Binary splitting method.}
This method is used to calculate the values of series with rational
coefficients, in particular, to calculate the hypergeometric series of
the form
\begin{equation*}
S=\sum_{i=0}^{\infty}{\frac{a(i)}{b(i)}
\prod_{j=0}^{i}{\frac{p(j)}{q(j)}}},
\end{equation*}
where $a$, $b$, $p$, and $q$ are polynomials with integer coefficients.
Linearly convergent hypergeometric series are used to calculate
many constants of analysis and elementary functions at
rational points; this series is linearly convergent if its partial sum
\begin{align}
S(\mu(k))=\sum_{i=0}^{\mu(k)}{\frac{a(i)}{b(i)}
\prod_{j=0}^{i}{\frac{p(j)}{q(j)}}},
\label{Eq:SeriesPartialSum}
\end{align}
where $\mu(k)$ is a linear function of $k$, differs from the exact value
by not more than $2^{-k}$:
\begin{align*}
|S-S(\mu(k))|\le 2^{-k}.
\end{align*}
In its classical variant, the binary splitting method works as follows.
Put $k_1=\mu(k)$. We consider the partial sum \eqref{Eq:SeriesPartialSum} for some
integers $i_1$ and $i_2$, $0\le i_1\le k_1$, $0\le i_2\le k_1$,
$i_1\le i_2$:
\begin{align*}
S(i_1,i_2)=\sum_{i=i_1}^{i_2}{\frac{a(i)p(i_1)
\ldots p(i)}{b(i)q(i_1)\ldots q(i)}}.
\end{align*}
We calculate $P(i_1,i_2)=p(i_1)\ldots p(i_2)$,
$Q(i_1,i_2)=q(i_1)\ldots q(i_2)$, $B(i_1,i_2)=b(i_1)\ldots b (i_2)$,
and $T(i_1,i_2)=B(i_1,i_2)Q(i_1,i_2)S(i_1,i_2)$. If $i_1=i_2$ then these values are
calculated directly. Otherwise, the series is divided into two parts, left and
right, and $P(i_1,i_2)$, $Q(i_1,i_2)$, and $B(i_1,i_2)$ are calculated
for each part recursively. Then the values obtained are combined:
\begin{align}
\begin{split}
P(i_1,i_2)&=P_l P_r, \quad Q(i_1,i_2)=Q_l Q_r, \quad B(i_1,i_2)=B_l B_r,\\
T(i_1,i_2)&=B_r Q_r T_l + B_l P_l T_r.
\end{split}
\label{Eq:BinSplitRecScheme}
\end{align}
The algorithm starts with $i_1=0$, $i_2=k_1$. After calculating
$T(0,k_1)$, $B(0,k_1)$, and $Q(0,k_1)$, we divide $T(0,k_1)$ by
$B(0,k_1)Q(0,k_1)$ to get the result with the given
accuracy. We write out the binary splitting method explicitly.

\algdef
{{\normalfont\itshape BinSplit}}
{Approximate value of the partial sum
\eqref{Eq:SeriesPartialSum} with accuracy $2^{-k}$}
{Record of accuracy $2^{-k}$}
{Approximate value of \eqref{Eq:SeriesPartialSum} with accuracy $2^{-k}$}
\begin {enumerate}
\item[1)]{$k_1:=\mu(k)$;}
\item[2)]{$[P,Q,B,T]:=BinSplitRecurs(0,k_1)$;}
\item[3)]{perform division $r:=\frac{T}{BQ}$ with accuracy $2^{-k}$;}
\item[4)]{return result $r$.}
\end {enumerate}

This algorithm uses the following subalgorithm computing recursively the values
$P$, $Q$, $B$, and $T$.

\algdef
{{\normalfont\itshape BinSplitRecurs}}
{Calculation of $P$, $Q$, $B$, and $T$}
{Bounds $i_1$, $i_2$ of the interval}
{Tuple [$P(i_1,i_2)$, $Q(i_1,i_2)$, $B(i_1,i_2)$, $T(i_1,i_2)$]}
\begin{enumerate}
\item[1)]{if $i_1=i_2$ then $T:=a(i_1)p(i_1)$ and return
tuple $[p(i_1),q(i_1),b(i_1),T]$;}
\item[2)]{$i_{mid}:=\frac{i_1+i_2}{2}$;}
\item[3)]{calculate $[P_l,Q_l,B_l,T_l]:=BinSplitRecurs(i_1,i_{mid})$;}
\item[4)]{calculate $[P_r,Q_r,B_r,T_r]:=BinSplitRecurs(i_{mid},i_2)$;}
\item[5)]{$T:=B_r Q_r T_l+B_l P_l T_r$ and return tuple
$[P_lP_r,Q_lQ_r,B_lB_r,T]$.}
\end{enumerate}

The lengths of $T(0,k_1)$ and $B(0,k_1)Q(0,k_1)$ are proportional to
$k\log(k)$; therefore the binary splitting method is quasi-linear
in space; the time complexity of this algorithm is $O(M(k)\log(k)^2)$
\cite{Haible98}.


\paragraph{5. Karatsuba's method for fast evaluation of $\exp(x)$.}
We consider the Taylor series of the real exponential function
\begin{equation}
\exp(x)=\sum_{i=0}^{\infty}{\frac{x^i}{i!}}=
1+\frac{x}{1!}+\frac{x^2}{2!}+\ldots+\frac{x^n}{n!}+\ldots
\label{Eq:ExpTeylorSeries}
\end{equation}
at $x_0$, $-\frac{1}{4}+2^{-m}<x_0<\frac{1}{4}-2^{-m}$.
We compute the value of this series with accuracy $2^{-n_3}$,
$n_3>7$, at the dyadic rational point $x_{m}$, $|x_{m}-x_0|\le 2^{-m}$,
$-\frac{1}{4}<x_{m}<\frac{1}{4}$. Let $k$ be the smallest value such that
\begin{equation}
n_3+1\le 2^{k},\quad\text{ }\quad m=2^{k+1}.
\label{Eq:ConstraintsForn3m}
\end{equation}
We represent
$x_{m}=\pm 0.00\alpha_3\alpha_4\ldots\alpha_{m}\alpha_{m+1}$ as
\begin{align*}
x_{m}&=
\pm 0.00\alpha_3\alpha_4 +
\pm 0.0000\alpha_5\alpha_6\alpha_7\alpha_8 + \ldots +
\pm 0.00\ldots 0 a_{m-2^k+1}a_{m-2^k+2}\ldots \alpha_{m}\alpha_{m+1}=\\
&=\frac{\beta_2}{2^4}+\frac{\beta_3}{2^8}+\frac{\beta_4}{2^{16}}+\ldots+
\frac{\beta_{k+1}}{2^{m}}=
\gamma_2+\gamma_3+\ldots+\gamma_{k+1},
\end{align*}
where $\beta_2=\pm \alpha_3\alpha_4$,
$\beta_3=\pm \alpha_5\alpha_6\alpha_7\alpha_8$, $\ldots$,
$\beta_{k+1}=\pm a_{m-2^k+1}a_{m-2^k+2}\ldots\alpha_{m}\alpha_{m+1}$;
$\gamma_{\nu}=\beta_{\nu} 2^{-2^{\nu}}$, $2\le\nu\le k+1$; $\beta_{\nu}$ is a
$2^{\nu-1}$-digit number.
We write $\exp(x_{m})$ as a product:
\begin{align}
\exp(x_{m})&=\exp(\gamma_2)\exp(\gamma_3)\ldots\exp(\gamma_{k+1}).
\label{Eq:ExpAsProductOfExpGamma}
\end{align}
In Karatsuba's method (FEE method, fast evaluation of the exponent; this
method is also known as the Brent trick [5]) the values $\exp(\gamma_{\nu})$
are then calculated using a Taylor series \eqref{Eq:ExpTeylorSeries}:
\begin{equation}
\exp(\gamma_{\nu})=
1+\frac{\beta_{\nu}}{1!2^{2^{\nu}}}+
\frac{\beta_{\nu}^2}{2!2^{2\cdot 2^{\nu}}}+\ldots+
\frac{\beta_{\nu}^r}{r!2^{r\cdot 2^{\nu}}}+R_{\nu}(r)=
\xi_{\nu}+R_{\nu}(r);
\label{Eq:ExpGammaNu}
\end{equation}
here $r=m 2^{-\nu+1}$. As for the residual sum, the following
inequality is satisfied \cite{Karatsuba91}:
\begin{equation*}
|R_{\nu}(r)|<2\frac{|\beta_{\nu}|^{r+1}}{(r+1)!\cdot 2^{(r+1)2^{\nu}}},
\end{equation*}
then $|R_{\nu}(r)|<2^{-m}$.
The values $\xi_{\nu}$ are obtained from the formulas
\begin{equation*}
\xi_{\nu}=\frac{a_{\nu}}{b_{\nu}},\quad
a_{\nu}=\xi_{\nu} b_{\nu},\quad
b_{\nu}=r! 2^{r 2^{\nu}},
\end{equation*}
where the integers $a_{\nu}$ are computed using a sequential process of
grouping members of the series \eqref{Eq:ExpGammaNu}. Note that the formula for $b_{\nu}$
contains the factorial of $r$, and $r$ varies from $m 2^{-1}$ to $m 2^{-k}$.
Hence there are values which are proportional to $n_3!$, and therefore
the length of the intermediate results is proportional
to $n_3\log(n_3)$.


\paragraph{6. Modification of the binary splitting method.}
We will modify the binary splitting method for the evaluation of
the hypergeometric series \eqref{Eq:ExpGammaNu} so that the calculations
are in the class \classname{Sch(FQLINTIME//LINSPACE)}.

We take $r=m 2^{-\nu+2}$, $k_1=\log(r)$, and $r_1=\lceil\frac{r}{k_1}\rceil$
($k_1$ is a natural number), and write the partial sum of
\eqref{Eq:ExpGammaNu} as
\begin{align}
P(\gamma_{\nu})
=\sigma_1+\tau_2[\sigma_2+\tau_3[\sigma_3+\ldots+
\tau_{k_1-1}[\sigma_{k_1-1}+\tau_{k_1}\sigma_{k_1}]]],
\label{Eq:PGammaNu}
\end{align}
where
\begin{align*}
\sigma_1&=
1+\frac{\beta_{\nu}}{1!2^{2^{\nu}}}+
\frac{\beta_{\nu}^2}{2!2^{2\cdot 2^{\nu}}}+\ldots+
\frac{\beta_{\nu}^{r_1-1}}{(r_1-1)!2^{(r_1-1)\cdot 2^{\nu}}},\\
\tau_2&=
\frac{\beta_{\nu}^{r_1}}{r_1!2^{r_1\cdot 2^{\nu}}},\\
\sigma_2&=
1+\frac{\beta_{\nu}}{(r_1+1)2^{2^{\nu}}}+
\frac{\beta_{\nu}^2}{(r_1+1)(r_1+2)2^{2\cdot 2^{\nu}}}+\ldots+
\frac{\beta_{\nu}^{r_1-1}}{(r_1+1)\ldots(2r_1-1)2^{(r_1-1)\cdot 2^{\nu}}},
\\
\tau_3&=
\frac{\beta_{\nu}^{r_1}}{(r_1+1)\ldots 2r_12^{r_1\cdot 2^{\nu}}},\\
\sigma_3&=
1+\frac{\beta_{\nu}}{(2r_1+1)2^{2^{\nu}}}+
\frac{\beta_{\nu}^2}{(2r_1+1)(2r_1+2)2^{2\cdot 2^{\nu}}}+\ldots+
\frac{\beta_{\nu}^{r_1-1}}{(2r_1+1)\ldots(3r_1-1)2^{(r_1-1)\cdot 2^{\nu}}}\\
&\cdots
\end{align*}
The $\sigma_t$ are calculated by the classical binary splitting method
for the sum \eqref{Eq:SeriesPartialSum}, where
\begin{align}
\begin{split}
\mu(k)&=r_1-1,\quad a(i)=1,\quad b(i)=1,\\
p(j)&=\begin{cases}1 & j=0\\ \beta_{\nu} & j\ne 0 \end{cases},\quad
q(j)=\begin{cases}1 & j=0\\ ((t-1)r_1+j)2^{2^{\nu}} & j\ne 0 \end{cases}.
\end{split}
\label{Eq:SigmaSeriesCoeffs}
\end{align}

We estimate the computational complexity of
the calculation of $\sigma_t$ and $\tau_t$ in the following two lemmas.

\begin{lemma}
\label{Lem:1}
The time complexity of the binary splitting method for the calculation
of $\sigma_t$ on the Schonhage machine is bounded above by $(r\log(r)+m)\log(m)\log\log(m)$;
the space complexity is bounded above by $O(m)$, where $m$ is given
by \eqref{Eq:ConstraintsForn3m}.
\begin{proof}
We consider an arbitrary maximal chain of recursive calls
derived from the calculations in accordance with algorithm
{\itshape BinSplitRecurs}. We consider pairs $(P,i)$, $(Q,i)$,
$(B,i)$, and $(T,i)$ where $i$ is the number of an element of the chain
of recursive calls and $P$, $Q$, $B$, and $T$ are in the $i$th element
of the chain. We suppose that the numbering in the chain begins with its
deepest element: $i=1,\ldots,\varsigma$, where $\varsigma$ is
the length of the chain, $\varsigma\le\lceil\log(2r_1)\rceil$.

Using mathematical induction on $i$, we show that the length of the
representation of $T$ in the pair $(T,i)$ satisfies $l(T)<2^{i}l(u)+2^{i}$,
where $u=r 2^{2^{\nu}}$. We note that $l(\beta_{\nu})<l(u)$
since $\beta_{\nu}$ is a $2^{\nu-1}$-digit integer. Next,
$l(P)<2^{i}l(u)$, $l(Q)<2^{i}l(u)$ for pairs $(P,i)$, $(Q,i)$,
since increasing $i$ leads to doubling the length;
$B$ is always $1$.

The induction begins with $i=1$: $l(T)\le l(u)<2^{1}l(u)+2^{1}$.
This inequality follows from \eqref{Eq:SigmaSeriesCoeffs} and the formula
$T=a(i_1)p(i_1)$ for $(T,1)$. Next, the induction step is
\begin{align*}
l(T)
<2^{i}l(u)+2^{i}l(u)+2^{i}+1
\le 2^{i+1}l(u)+2^{i+1}.
\end{align*}
This inequality follows from \eqref{Eq:BinSplitRecScheme}:
there are two multiplications of numbers of length $2^{i}l(u)$
and $2^{i}l(u)+2^{i}$ and one addition.

We note that, based on this inequality, $T$ of the pair $(T,\varsigma)$
has length $O(m)$ because
\begin{align*}
l(T)
&<C_1 2^{\varsigma}l(u)\le C_2\frac{r}{\log(r)}(\log(r)+2^{\nu})
\le C_2 \left[r + \frac{m 2^{-\nu+1}2^{\nu}}{\log(r)}\right]
\le C_3 m.
\end{align*}
We estimate the time complexity of the calculation of $\sigma_t$.
We must take into account the property of the complexity of integer
multiplication: $2M(2^{-1}m)\le M(m)$ (quasi-linear and polynomial functions
satisfy this property). Since at a tree node of recursive calls at
level $i$ there are $C_4$ multiplications of $2^{\varsigma-i}$ numbers
of length at most $2^{i}l(u)+2^{i}$ and $h\le C\log(m)$, we obtain
the following estimate for the number of operations required
to calculate $\sigma_t$:
\begin{align*}
Time(\sigma_t)
&\le C_4\sum_{i=1}^{\varsigma}{2^{\varsigma-i}M(2^{i}l(u)+2^{i})}
\le C_5\sum_{i=1}^{\varsigma}{2^{\varsigma-i}M(2^{i+1}l(u))}\\
&=C_5\sum_{i=1}^{\varsigma}{2^{\varsigma-i}M(2^{\varsigma-(\varsigma-(i+1))}l(u))}
\le C_6\sum_{i=1}^{\varsigma}{2^{\varsigma-i}2^{-\varsigma+(i+1)}M(2^{\varsigma}l(u))}\\
&\le C_7 \varsigma M(2^{\varsigma}l(u))
\le C_8\log(r)M\left(r+\frac{m}{\log(r)}\right)
\end{align*}
(here we use the inequality for $2^{\varsigma}l(u)$ from the estimate of
$l(T)$ for the pair $(T,\varsigma)$). The final division gives $O(M(m))$ operations.
If we use the Schonhage--Strassen algorithm for integer multiplication,
then
\begin{align}
\begin{split}
Time(\sigma_t)
&\le C_9\left(r+\frac{m}{\log(r)}\right)\log(r)\log(m)\log\log(m)=\\
&=C_9(r\log(r)+m)\log(m)\log\log(m).
\end{split}
\label{Eq:TimeSigmat}
\end{align}

Now we estimate the space complexity of the computation of $\sigma_t$.
At an element of a chain of recursive calls with number
$i$, the amount of memory consumed for the temporary variables is
$C_{10}(2^{i}l(u)+2^{i})$. Hence, we conclude that the amount of memory in
all the simultaneously existing recursive calls in the chain is
estimated as follows:
\begin{align*}
Space(\sigma_t)
\le\sum_{i=1}^{\varsigma}{C_{10}(2^{i}l(u)+2^{i})}
\le C_{11}(2^{\varsigma}l(u)+2^{\varsigma})
\le C_{12} m=O(m).
\end{align*}
\end{proof}
\end{lemma}

\begin{lemma}
\label{Lem:2}
The time complexity of the calculation of $\tau_t$ using binary splitting
method on a Schonhage machine is bounded above by $(r\log(r)+m)\log(m)\log\log(m)$;
the space complexity is bounded above by $O(m)$ where $m$ is given
by \eqref{Eq:ConstraintsForn3m}.
\begin{proof}
The estimates of the computational complexity of $\tau_t$ are the same as
those for $\sigma_t$ due to the fact that the inequalities in the proof of
Lemma \ref{Lem:1} are also suitable for $\tau_t$ (we can calculate the
numerator and denominator of $\sigma_t$ using the binary splitting method
for products).
\end{proof}
\end{lemma}

We calculate approximate values $P(\gamma_{\nu})^*$ with accuracy
$2^{-(m+1)}$ by \eqref{Eq:PGammaNu} using the
following iterative process:
\begin{align}
\begin{split}
&h_1(m_1)=\sigma_{k_1}^*, \\
&\widehat h_i(m_1)=\sigma_{k_1-i+1}^*+\tau_{k_1-i+2}^* h_{i-1},
\quad i=1,\dots,k_1,\\
&h_i(m_1)=\widehat h_i(m_1)+\varepsilon_i;
\end{split}
\label{Eq:PGammaNuhScheme}
\end{align}
for $i=k_1$ we set $P(\gamma_{\nu})^*= h_{k_1}(m_1)$. Here $m_1\ge m$
($m_1$ will be chosen later), and $\sigma_i^*$ and $\tau_i^*$ are approximations
of $\sigma_i$ and $\tau_i$ with accuracy $2^{-m_1}$.
The $h_i(m_1)$ are obtained by discarding bits
$q_{m_1+1}q_{m_1+2}\dots q_{m_1+j}$ of numbers $\widehat h_i(m_1)$ after
the binary point starting with the $(m_1+1)$th bit:
\begin{align}
|\varepsilon_i|=|h_i(m_1)-\widehat h_i(m_1)|=
0.0\dots 0 q_{m_1+1}q_{m_1+2}\dots q_{m_1+j},
\label{Eq:PGammaNuVarepsilon}
\end{align}
and the sign of $\varepsilon_i$ is the same as the sign of $\widehat h_i(m_1)$
(it is clear that $|\varepsilon_i|<2^{-m_1}$).

We note that the accuracy of $\exp(\gamma_{\nu})$ is $2^{-m}$ since
we compute $\xi_{\nu}^*=P(\gamma_{\nu})^*$ with accuracy $2^{-(m+1)}$
and
\begin{align*}
|\exp(\gamma_{\nu})^*-\exp(\gamma_{\nu})|\le
|\xi_{\nu}^*-\xi_{\nu}|+|R_{\nu}(r)|<
2^{-(m+1)}+2^{-(m+1)}=2^{-m}.
\end{align*}

\begin{lemma}
\label{Lem:3}
For every $i\in 1\dots k_1$
\begin{align}
|h_i(m_1)|<2.
\label{Eq:Lemma3Enequality}
\end{align}
\begin{proof}
We apply mathematical induction on $j$ for $h_j(m_1)$ using the estimates
\begin{align*}
|\sigma_i|<\exp(\gamma_{\nu})<\frac{4}{3},\quad
\tau_i<{\gamma_{\nu}}^{r_1}\le\frac{1}{4}.
\end{align*}
The induction base for $j$ is $j=1$: $|h_1(m_1)|\le\sigma_{k_1}+2^{-m_1}<2$.
The induction step for $(j+1)\geq 2$:
\begin{align*}
|h_{j+1}(m_1)|
&=|\sigma_{k_1-(j+1)+1}^* + \tau_{k_1-(j+1)+2}^* h_{j}+\varepsilon_{j+1}|\\
&\le\frac{4}{3}+\left[\frac{1}{4}+2^{-m_1}\right]2+2^{-m_1}<2.
\end{align*}
\end{proof}
\end{lemma}

\begin{lemma}
\label{Lem:Lemma4}
The error of the calculation of $h_{k_1}(m_1)$ using the scheme
{\normalfont\eqref{Eq:PGammaNuhScheme}} is estimated to be
\begin{align*}
\Delta(k_1,m_1)<2^{-m_1+k_1}.
\end{align*}
\begin{proof}
We put
\begin{align*}
H_1=\sigma_{k_1},\quad
H_i=\sigma_{k_1-i+1} + \tau_{k_1-i+2}H_{i-1},\quad
\eta(i,m_1)=|h_i(m_1)-H_i|.
\end{align*}
We use mathematical induction for $\eta(j,m_1)$ on $j$.
The induction base for $j$ is $1$:
\begin{align*}
\eta(1,m_1)=|h_1(m_1)-H_1|=|\sigma_{k_1}^*-\sigma_{k_1}|<2^{-m_1+1}.
\end{align*}
The induction step is $(j+1)\geq 2$:
\begin{align*}
\eta(j+1,m_1)
&=|\sigma_{k_1-(j+)+1}^* + \tau_{k_1-(j+1)+2}^* h_{j}(m_1)+\varepsilon_{j+1}-
\sigma_{k_1-(j+1)+1} - \tau_{k_1-(j+1)+2}H_{j}|\\
&<|\tau_{\upsilon}^* h_{j}(m_1)-\tau_{\upsilon}h_{j}(m_1)+
\tau_{\upsilon}h_{j}(m_1)-\tau_{\upsilon}H_{j}|+2\cdot 2^{-m_1}\\
&\leq 2^{-m_1}h_{j}(m_1)+2^{-2}\eta(j,m_1)+2\cdot 2^{-m_1}.
\end{align*}
Since \eqref{Eq:Lemma3Enequality}, $|h_j(m_1)|<2$.  By the induction
hypothesis, $\eta(j,m_1)<2^{-m_1+j}$, and so we get
\begin{align*}
\eta(j+1,m_1)
<2\cdot 2^{-m_1}+2^{-2}2^{-m_1+j}+2\cdot 2^{-m_1}<2^{-m_1+(j+1)}.
\end{align*}
From $\Delta(k_1,m_1)=\eta(k_1,m_1)$ we now obtain the required inequality.
\end{proof}
\end{lemma}

Lemma \ref{Lem:Lemma4} implies that it is sufficient to take $m_1=2m+1$
to compute $P(\gamma_{\nu})$ with an accuracy of $2^{-(m+1)}$.

We denote the algorithm for the calculation of the hypergeometric series
using scheme \eqref{Eq:PGammaNuhScheme} by \algname{RLinSpaceBinSplit}
(linear space binary splitting).

\algdef
{\algname{RLinSpaceBinSplit}}
{The approximate value of the hypergeometric series}
{Record of the accuracy $2^{-m}$}
{The approximate value of \eqref{Eq:ExpGammaNu} with accuracy $2^{-m}$}
\begin{enumerate}
\item[1)]{$m_1:=2m+1$;}
\item[2)]{$h:=\sigma_{k_1}^*$ (using the classical binary splitting method
with accuracy $2^{-m_1}$);}
\item[3)]{make a loop through $i$ from $2$ to $k_1$:
  \begin{enumerate}
  \item[a)]{calculate $v_1:=\sigma_{k_1-i+1}^*$ with accuracy $2^{-m_1}$
  using the classical binary splitting method and $v_2:=\tau_{k_1-i+2}^*$ with
  accuracy $2^{-m_1}$,}
  \item[b)]{calculate $\widehat h:=v_1+v_2 h$,}
  \item[c)]{assign value $\widehat h$ to $h$ rounded
  in accordance with \eqref{Eq:PGammaNuVarepsilon};}
  \end{enumerate}
}
\item[4)]{write $h$ on exit.}
\end{enumerate}

We estimate the time computational complexity of this algorithm on
a Schonhage machine, taking into account that $m$ dependends linearly on $n_3$:
\begin{itemize}
\item{$\log(r)$ computations of $\sigma_t$ give the following (from
inequality \eqref{Eq:TimeSigmat}):
\begin{align*}
Time(all(\sigma_t))
&\le\sum^{\log(m)}_{\nu=2}{C_9(r\log(r)+m)\log(m)\log\log(m)}\le\\
&\le C_{13}(m\log(m)^2\log\log(m))=O(M(n_3)\log(n_3));
\end{align*}
}
\item{$O(\log(n_3))$ computations of $\tau_t$ give
$O(M(n_3)\log(n_3))$;}
\item{$O(\log(n_3))$ multiplications of numbers of the length $O(n_3)$ give
$O(M(n_3)\log(n_3))$;}
\end{itemize}
in total we obtain $O(M(n_3)\log(n_3))$. The space complexity
of the modified binary splitiing method \algname{RLinSpaceBinSplit}
is $O(n_3)$ since in all calculations in this algorithm we process
numbers of length $O(n_3)$.

\begin{proposition}
The modified binary splitting algorithm \algnameindef{RLinSpaceBinSplit}
belongs to the class \classnameindef{Sch(FQLINTIME//LINSPACE)}.
\end{proposition}


\paragraph{7. Modification of FEE.}
We construct a modification of the method FEE so that our calculations
are in class $Sch(FQLINTIME//LINSPACE)$.

In the classical algorithm FEE the values $\exp(x_{m })^*$ are calculated
by \eqref{Eq:ExpAsProductOfExpGamma} using a pairwise summation of
the numbers $\exp(\gamma_i)^*$; in FEE we need to keep in memory
$O(\log(\log(n_3)))$ numbers of length $O(n_3)$ and, as already mentioned,
$n_3!$ values are processed in this algorithm, so it does not have linear space complexity.

We calculate $\exp(x_{m })^*$ with accuracy $2^{-n_3}$ using the following iterative process:
\begin{align}
\begin{split}
&h_2(m)=\exp(\gamma_2)^*, \\
&\widehat h_i(m)=h_{i-1}(m)\exp(\gamma_i)^*,
\quad i=2,\dots,k+1,\\
&h_i(m)=\widehat h_i(m)+\varepsilon_i;
\end{split}
\label{Eq:FEEhSchema}
\end{align}
for $i=k+1$ we put $\exp(x_{m})^*=h_{k+1}(m)$. Here $\exp(\gamma_i)^*$
are approximations for the values $\exp(\gamma_i)$ with accuracy $2^{-m}$
obtained from formula \eqref{Eq:ExpGammaNu}.
The values $h_i(m)$ are obtained by discarding bits
$q_{m+1}q_{m+2}\dots q_{m+t}$ of the numbers $\widehat h_i(m)$ after
the binary point starting with the $(m+1)$th bit, i.e.,
\begin{align}
|\varepsilon_i|=|h_i(m)-\widehat h_i(m)|=
0.0\dots 0 q_{m+1}q_{m+2}\dots q_{m+t},
\label{Eq:FEEvarepsilon}
\end{align}
and the sign of $\varepsilon_i$ is the same as the sign of $\widehat h_i(m)$
(it is clear that $|\varepsilon_i|<2^{-m}$).

\begin{lemma}
\label{Lem:5}
For every $i\in 2\dots k+1$
\begin{align}
|h_i(m)|<2^{i-1}.
\label{Eq:Lemma5Enequality}
\end{align}
\begin{proof}
We apply mathematical induction on $j$ for $h_j(m)$.
The induction base is $j=2$:
\begin{align*}
|h_2(m)|\leq|\exp(\gamma_2)|+2^{-m}<\frac{3}{2}+2^{-16}<2^{2-1}
\end{align*}
(here we take into account that $|\gamma_i|<\frac{1}{4}$, $m\ge 16$).
The induction step is $(j+1)\geq 3$:
\begin{align*}
|h_{j+1}(m)|
&=|h_{j}(m)\exp(\gamma_{j+1})^*+\varepsilon_{j+1}|
<h_{j}(m)(\exp(\gamma_{j+1})+2^{-m})+2^{-m}\\
&<2^{j-1}\left[\frac{3}{2}+2^{-m}\right]+2^{-m}<2^{j}.
\end{align*}
\end{proof}
\end{lemma}

\begin{lemma}
\label{Lem:6}
The error of the calculation of $h_{k+1}(m)$ using scheme
{\normalfont\eqref{Eq:FEEhSchema}} is estimated to be
\begin{align*}
\Delta(k+1,m)<2^{-m+2(k+1)}.
\end{align*}
\begin{proof}
We put
\begin{align*}
H_2=\exp(\gamma_2),\quad
H_i=\exp(\gamma_2)\exp(\gamma_3)\ldots\exp(\gamma_{i}),\quad
\eta(i,m)=|h_i(m)-H_i|.
\end{align*}
We use mathematical induction for $\eta(j,m)$ on $j$. The induction base
is $j=2$:
\begin{align*}
\eta(2,m)=|h_2(m)-H_2|=|\exp({\gamma_2})^*-\exp({\gamma_2})|\le 2^{-m}
<2^{-m+2(1+1)}.
\end{align*}
The induction step is $(j+1)\geq 3$:
\begin{align*}
\eta(j+1,m)
&=|h_{j}(m)\exp({\gamma_{j+1}})^*+\varepsilon_{j+1}-
H_{j}\exp({\gamma_{j+1}})|\\ 
&<|h_{j}(m)\exp({\gamma_{j+1}})^*-h_{j}(m)\exp({\gamma_{j+1}})+\\
&\qquad h_{j}(m)\exp({\gamma_{j+1}})-H_{j}\exp({\gamma_{j+1}})|+2^{-m}\\
&\leq h_{j}(m)2^{-m}+\eta(j,m)\exp({\gamma_{j+1}})+2^{-m}.
\end{align*}
Since \eqref{Eq:Lemma5Enequality}, $|h_j(m)|<2^{j-1}$. Then by the induction
hypothesis, $\eta(j,m)<2^{-m+2j}$, and so we get
\begin{align*}
\eta(j+1,m)
<2^{j-1}2^{-m}+\left[2^{-m+2j}\right]\frac{3}{2}<2^{-m+2(j+1)}.
\end{align*}
From $\Delta(k+1,m)=\eta(k+1,m)$, we now get the required inequality.
\end{proof}
\end{lemma}

Lemma \ref{Lem:6} implies that the accuracy $2^{-m}$ of the calculation of
$\exp({\gamma_i})^*$ is sufficient to calculate $\exp(x_m)^*$
with accuracy $2^{-n_3}$, since
\begin{align*}
&-m+2(k+1)\le -n_3\thickspace\Rightarrow
\thickspace 2^{k+1}-2(k+1)\ge n_3\quad \text{and}\\
&2^{k+1}-2(k+1)>2^{k+1}2^{-1}=2^{k}\ge n_3+1
\end{align*}
(here we recall that $n_3+1\le 2^{k}$).

We denote the algorithm of the calculation of the real exponential function
using scheme \eqref{Eq:FEEhSchema} by \algname{RLinSpaceFEE}
(linear space fast exponential evaluation).

\algdefo
{\algname{RLinSpaceFEE}}
{The approximate value of the real exponential function on the interval
$\left[-\frac{1}{8},\frac{1}{8}\right]$}
{Record of the accuracy $2^{-n_3}$}
{The approximate value $\exp(x)$ with accuracy $2^{-n_3}$}
{$\phi_x$}
\begin{enumerate}
\item[1)]{$h:=\exp(\gamma_2)^*$ (using algorithm \algname{RLinSpaceBinSplit}
with accuracy $2^{-m}$);}
\item[2)]{make a loop through $i$ from $3$ to $k+1$:
  \begin{enumerate}
  \item[a)]{calculate $v_1:=\exp(\gamma_i)^*$ using algorithm
  \algname{RLinSpaceBinSplit} with accuracy $2^{-m}$,}
  \item[b)]{calculate $\widehat h:=h\cdot v_1$,}
  \item[c)]{assign the value $\widehat h$ to $h$ rounded
  in accordance with \eqref{Eq:FEEvarepsilon};}
  \end{enumerate}
}
\item[3)]{write $h$ on exit.}
\end{enumerate}                                    

The 
time complexity of this algorithm on a Schonhage is $O(M(n_3)\log(n_3)^2)$
as the algorithm of the calculation of the hypergeometric series
\algname{RLinSpaceBinSplit} uses $O(M(n_3)\log(n_3))$ operations,
and in scheme \eqref{Eq:FEEhSchema} there are $O(\log(n_3))$ such
calculations and $O(\log(n_3))$ multiplications of numbers of the
length $O(n_3)$; the space complexity of \algname{RLinSpaceFEE} is
$O(n_3)$ since in all the calculations in this algorithm numbers of
length $O(n_3)$ are used.

\begin{proposition}
The modified FEE algorithm \algname{RLinSpaceFEE} for the calculation
of the exponential function belongs to the class
\classname{Sch(FQLINTIME//LINSPACE)}.
\end{proposition}


\paragraph{9. Calculation of the real function $\exp(x)$.}
Let $p$ be a positive integer, $p\geq 0$.
We compute the function $\exp(x)$ with accuracy $2^{-n}$ in
the interval $[-2^{p},2^{p}]$.

We perform the multiplicative reduction of the interval of the complex argument.
Namely, we take an integer $s=2^{p+3}$ and $x'=\frac{x}{s}$; then
$|x'|=\frac{|x|}{s}$, i.e., $x'$ is in the interval $[-2^{-3}\le x'\le 2^{-3}]$.
Thus the calculation of $\exp(x)$ is reduced to the computation of $\exp(x')$
and then we raise this value to the power $s$ to get the result. It
is easy to see that the dependency function for $n_1$ of the accuracy
of $\exp(x')$ is $n_1=L(n)+C(p)$, where $L(n)$ is a linear function
of $n$, and $C(p)$ is a constant which is independent of $p$
(constant in the sense that it doesn't depend on $n$).

Put $m\ge n_1+3$. We then have: $x_{m}=x_{0}+\theta_1 2^{-m}$,
$|\theta_1|\le 1$; $|x_0|\le 2^{-3}$, and
\begin{align*}
|\exp(x_0)-\exp(x_{m})|
&=
|\exp(x_0)-\exp(x_0)\exp(\theta_1 2^{-m})|=
\exp(x_0)|1-\exp(\theta_1 2^{-m})|.
\end{align*}
Since $\exp(\theta_1 2^{-m})<\frac{1}{1-2^{-m}}$,
$|\exp(\theta_1 2^{-m})-1|<\frac{2^{-m}}{1-2^{-m}}<2^{-m+1}$.
Therefore we have the estimate
\begin{align*}
|\exp(x_0)-\exp(x_{m})|<2\cdot 2^{-m+1}=2^{-m+2}\le 2^{-(n_1+1)},
\end{align*}
which shows provided the accuracy of the calculation of $x_{m}$ is better than
$2^{-(n_1+3)}$, then one achieves an accuracy $2^{-n_2}$, $n_2=n_1+1$ for
$\exp(x_0)$. Since $m=2^{k+1}$, this condition is satisfied.

Now we need to keep in mind that we calculate the approximate value of
$\exp(x_{m})^*$. If this approximation is calculated with accuracy
$2^{-n_3}$, $n_3=n_1+1$, then
\begin{align*}
|\exp(x_0)-\exp(x_{m})^*|
&\le|\exp(x_0)-\exp(x_{m})|+|\exp(x_{m})-\exp(x_{m})^*|\\
&<2^{-(n_1+1)}+2^{-(n_1+1)}=2^{-n_1}.
\end{align*}
This implies that we can take $n_3=n_1+1$ in algorithm \algname{RLinSpaceFEE}.

We are now ready to describe the algorithm.

\algdefpo
{\algname{RLinSpaceExpValue}}
{The approximate value of the complex exponential function}
{Record of the accuracy $2^{-n}$}
{The approximate value $\exp(z)$ with accuracy $2^{-n}$}
{$\phi_x$ for the argument $x$}
{Constant $p$}
\begin{enumerate}
\item[1)]{$n_1:=L(n)+C(p)$;}
\item[2)]{$n_3:=n_1+1$;}
\item[3)]{calculate $k$, $m$ so that \eqref{Eq:ConstraintsForn3m} holds;}
\item[4)]{$p_1:=p+3$}
\item[5)]{$s:=2^{p_1}$;}
\item[6)]{compute $x^*:=\phi_x(\max(1,m-p_1))$;}
\item[7)]{perform the reduction of the interval: $(x^*)'=\frac{x^*}{s}$
(the accuracy of the arguments will be $2^{-m}$);}
\item[8)]{using algorithm \algname{RLinSpaceFEE}, calculate $v:=\exp(x^*)^*$
with accuracy $2^{-n_3}$;}
\item[9)]{write complex number $v^s$ to the output.}
\end{enumerate}
The properties of algorithms \algname{TLinSpaceBinSplit} and 
\algname{TLinSpaceFEE} allow us assert
the following propositions.

\begin{proposition}
Algorithm \algname{RLinSpaceExpValue} of the calculation of the complex
exponential function belongs to the class
\classnameindef{Sch(FQLINTIME//LINSPACE)}.
\end{proposition}

The estimates of the computational complexity of algorithm
\algname{RLinSpaceExpValue} on the Schonhage machine are the same as those
for algorithm \algname{RLinSpaceFEE}: that is, the time complexity
is $O(M(n_3)\log(n_3)^2)$ and the space complexity is $O(n_3)$.
If we use the Schonhage--Strassen algorithm for integer multiplication,
then the time complexity of algorithm \algname{RLinSpaceExpValue}
is bounded above by $O(n_3\log(n_3)^3\log\log(n_3))$.

\begin{proposition}
The real function $\exp(x)$ is a \classname{Sch(FQLINTIME//LINSPACE)}
constructive real function in any interval $[-2^{p},2^{p}]$.
\end{proposition}

%
\paragraph{8. Calculation of the function $\exp(\mathbf{i}\cdot y)$.}
It is easy to show that all the algorithms, lemmas, and estimates can be
formulated for the evaluation of the function $\exp(\mathbf{i}\cdot y)$.
\begin{enumerate}
\item{Calculate \eqref{Eq:SeriesPartialSum}, where
$p(j)=\mathbf{i}\cdot p_{real}(j)$; $P$, $Q$, $B$ are integers, and $T$ is
complex. The estimates of the computational complexity are the same as
the estimates in Lemmas \ref{Lem:1} and \ref{Lem:2}. The $h_i(m)$ are obtained
by discarding bits $q_{m_1+2}q_{m_1+3}\dots q_{m_1+j}$ of the numbers
$\widehat h_i(m_1)$ after the binary point starting with the $(m_1+2)$th bit.
Lemmas \ref{Lem:3} and \ref{Lem:4} are true for the new scheme.
Algorithm \algname{CLinSpaceBinSplit} is the same as \algname{RLinSpaceFEE}.
}
\item{In the FEE method, we calculate
\begin{align*}
\exp(\mathbf{i}\cdot x_{m})
&=\exp(\mathbf{i}\cdot\gamma_2)\exp(\mathbf{i}\cdot\gamma_3)\ldots\exp(\mathbf{i}\cdot\gamma_{k+1}).
\end{align*}
The estimate for $|R_{\nu}(r)|$ is the same as that for $\exp(x)$,
that is, the series $Eq:RealExp:GammaNuSeries$ converges linearly for
$\exp(\mathbf{i}\cdot y)$.
}
\item{
In the formulas for $\sigma_i$ and $\tau_i$, we use $\mathbf{i}\cdot\beta_{\nu}$
and we use algorithm \algname{CLinSpaceBinSplit} for the computation
of $P(\mathbf{i}\cdot\gamma_{\nu})^*$.
}
\item{
In the scheme for the computation of $P(\mathbf{i}\cdot\gamma_{\nu})^*$,
the $h_i(m)$ are obtained by discarding bits $q_{m_1+2}q_{m_1+3}\dots q_{m_1+j}$
of the numbers $\widehat h_i(m_1)$ after the binary point starting with
the $(m_1+2)$th bit; Lemmas \ref{Lem:5} and \ref{Lem:5} are true for the new scheme;
algorithm \algname{CLinSpaceFEE} is the same as \algname{RLinSpaceFEE}.
}
\end{enumerate}
Denote the algorithm for the computation of $\exp(\mathbf{i}\cdot y)$
by \algname{CLinSpaceExpValue}.


\paragraph{9. Calculation of the complex function $\exp(z)$.}
Let $p$ be a positive integer, $p\geq 0$.
We compute the function $\exp(z)$ with accuracy $2^{-n}$ in
the area $|z|\le 2^p$ (we have
$|x|\le 2^p$, $|y|\le 2^p$).

We perform the multiplicative reduction of the interval of the complex argument.
Namely, we take an integer $s=2^{p+3}$ and $z'=\frac{z}{s}$; then
$|z'|=\frac{|z|}{s}$, i.e., $z'$ is in the area $|z'|\le 2^{-3}$ and
both $|x|\le 2^{-3}$ and $|y|\le 2^{-3}$. Thus
the calculation of $\exp(z)$ is reduced to the computation of $\exp(z')$
and then we raise this value to the power $s$ to get the result. It
is easy to see that the dependency function for $n_1$ of the accuracy
of $\exp(z')$ is $n_1=L(n)+C(p)$, where $L(n)$ is a linear function
of $n$, and $C(p)$ is a constant which is independent of $p$
(constant in the sense that it doesn't depend on $n$).

Next we consider the function $\exp(z)$ in the area $|z|\le 2^{-3}$.
Put $\zeta_x=\exp(x)$ and $\zeta_y=\exp(\mathbf{i}\cdot y)$.
According to \eqref{Eq:ComplNumberApprox}, we need to calculate 
$\zeta_x$ and $\zeta_y$ with accuracies of $2^{-n_2}$, $n_2=n_1+1$ respectively,
in order to calculate $\exp(z)$ with an accuracy of $2^{-n_1+1}$.

Put $m\ge n_1+3$. We have the following: $x_{m}=x_{0}+\theta_1 2^{-m}$,
$|\theta_1|\le 1$; at that $|x_0|\le 2^{-3}$ and
\begin{align*}
|\exp(x_0)-\exp(x_{m})|
&=
|\exp(x_0)-\exp(x_0)\exp(\theta_1 2^{-m})|=
\exp(x_0)|1-\exp(\theta_1 2^{-m})|.
\end{align*}
Since $\exp(\theta_1 2^{-m})<\frac{1}{1-2^{-m}}$,
$|\exp(\theta_1 2^{-m})-1|<\frac{2^{-m}}{1-2^{-m}}<2^{-m+1}$,
and therefore we have the estimate
\begin{align*}
|\exp(x_0)-\exp(x_{m})|<2\cdot 2^{-m+1}=2^{-m+2}\le 2^{-(n_1+1)},
\end{align*}
which shows that if accuracy of the calculation of $x_{m}$ is better than 
$2^{-(n_1+3)}$, then one achieves an accuracy of $2^{-(n_1+1)}$ for
$\exp(x_0)$. Since $m=2^{k+1}$, this condition is satisfied.
A similar estimate can be obtained for
$|\exp(\mathbf{i}\cdot y_0)-\exp(\mathbf{i}\cdot y_{m})|$.

Now we need to keep in mind that we calculate the approximate value of
$\exp(z_{m})^*$. If this approximation is calculated with accuracy
$2^{-n_3}$, $n_3=n_1+1$, then
\begin{align*}
|\exp(z_0)-\exp(z_{m})^*|
&\le|\exp(z_0)-\exp(z_{m})|+|\exp(z_{m})-\exp(z_{m})^*|\\
&<2^{-(n_1+1)}+2^{-(n_1+1)}=2^{-n_1}.
\end{align*}
This implies that we can take $n_3=n_1+1$ in algorithms \algname{RLinSpaceFEE} and
\algname{CLinSpaceFEE}.

We are now ready to describe the basic algorithm.

\algdefpo
{\algname{LinSpaceExpValue}}
{The approximate value of the complex exponential function}
{Record of the accuracy $2^{-n}$}
{The approximate value $\exp(z)$ with accuracy $2^{-n}$}
{$\phi_x$ and $\phi_y$ for the argument $z=x+i y$}
{Constant $p$}
\begin{enumerate}
\item[1)]{$n_1:=L(n)+C(p)$;}
\item[2)]{$n_3:=n_1+1$;}
\item[3)]{calculate $k$, $m$ so that \eqref{Eq:ConstraintsForn3m} holds;}
\item[4)]{$p_1:=p+3$}
\item[5)]{$s:=2^{p_1}$;}
\item[6)]{compute $x^*:=\phi_x(\max(1,m-p_1))$,
$y^*:=\phi_y(\max(1,m-p_1))$;}
\item[7)]{perform the reduction of the interval: $(x^*)'=\frac{x^*}{s}$,
$(y^*)'=\frac{y^*}{s}$ (the accuracy of the arguments will be $2^{-m}$);}
\item[8)]{using algorithm \algname{RLinSpaceFEE}, calculate $v_1:=\zeta_x^*$
with accuracy $2^{-n_3}$; using algorithm \algname{CLinSpaceFEE},
calculate $v_2:=\zeta_y^*$; here the arguments are $(x^*)'$, $(y^*)'$;}
\item[9)]{write to the output the complex number $(v_1+i\cdot v_2)^s$.}
\end{enumerate}
The properties of algorithms \algname{RLinSpaceBinSplit},
\algname{RLinSpaceFEE}, \algname{CLinSpaceBinSplit}, and
\algname{CLinSpaceFEE} allow us to assert
the following propositions.

\begin{proposition}
Algorithm \algname{CLinSpaceExpValue} of the calculation of the complex
exponential function belongs to the class
\classnameindef{Sch(FQLINTIME//LINSPACE)}.
\end{proposition}

The estimates of the computational complexity of algorithm
\algname{CLinSpaceExpValue} on a Schonhage machine are the same as those
of algorithms \algname{RLinSpaceFEE} and \algname{CLinSpaceFEE};
that is, the time complexity is $O(M(n_3)\log(n_3)^2)$ and the space
complexity is $O(n_3)$. If we use the Schonhage--Strassen algorithm
for integer multiplication, then the time complexity of algorithm
\algname{CLinSpaceExpValue} is bounded above by
$O(n_3\log(n_3)^3\log\log(n_3))$.

\begin{theorem}
The complex function $\exp(z)$ is a \classname{Sch(FQLINTIME//LINSPACE)}
constructive complex function in any area $|z|\le 2^{p}$.
\end{theorem}


\paragraph{10. Computation of the complex functions $\sin(z)$, $\cos(z)$,
$\sh(z)$, $\ch(z)$.}
Based on the formulas for the trigonometric functions
\begin{align*}
\sin(z)=\frac{e^{iz}-e^{-iz}}{2i}=i\frac{e^{iz}-e^{-iz}}{-2}, \quad
\cos(z)=\frac{e^{iz}+e^{-iz}}{2},
\end{align*}
we obtain the following:

\begin{proposition}
The complex function $\sin(z)$ is a \classnameindef{Sch(FQLINTIME//LINSPACE)}
constructive complex function in any area $|z|\le 2^{p}$.
\end{proposition}

\begin{proposition}
The complex function $\cos(z)$ is a \classnameindef{Sch(FQLINTIME//LINSPACE)}
constructive complex function in any area $|z|\le 2^{p}$.
\end{proposition}

The following two propositions also follow directly from the formulas
for the hyperbolic sine and cosine:
\begin{align*}
\sh(z)=\frac{e^{z}-e^{-z}}{2}, \quad \ch(x)=\frac{e^{z}+e^{-z}}{2}.
\end{align*}

\begin{proposition}
The complex function $\sh(z)$ is a \classnameindef{Sch(FQLINTIME//LINSPACE)}
constructive complex function in any area $|z|\le 2^{p}$.
\end{proposition}

\begin{proposition}
The complex function $\ch(z)$ is a \classnameindef{Sch(FQLINTIME//LINSPACE)}
constructive complex function in any area $|z|\le 2^{p}$.
\end{proposition}


\paragraph{11. Conclusion.}
Constructed algorithm \algname{CLinSpaceExpValue} can be used in computer
science as the basis of the \classname{Sch(FQLINTIME//LINSPACE)} constructive
complex functions $\exp(z)$, $\sin(z)$, $\cos(z)$, $\sh(z)$, $\ch(z)$,
defined on the set of \classname{Sch(FQLINTI\-ME//LINSPACE)} constructive
complex numbers.

Note also that if we use a simple recursive method for integer
multiplication with time complexity $O(n^{\log(3)})$, then the time
complexity of algorithm \algname{CLinSpaceExpValue} is
$O(n^{\log(3)}\log(n)^2)$.

As future research plans, we could note the problem of the construction of
algorithms based on series expansions for the \classname{Sch(FQLINTIME//LINSPACE)}
computable analogues of other elementary functions as well as the importance
of the probably more difficult problem of the construction of computable
analogues of elementary functions (also based on algorithms that use
series expansions) with a time complexity of $O(n\log(n)^k)$, $k\le 3$,
and with linear space complexity.


\end{document}